\documentstyle[twocolumn,aps,prl,epsf,epsfig]{revtex}

\begin{document} 
\twocolumn[\hsize\textwidth\columnwidth\hsize 
\csname@twocolumnfalse%
\endcsname 
\draft 
\title{Multipartite transmission of quantum solitons} 
\author{I.E.~Mazets$^1$\,$^,$\,$^2$, G.~Kurizki$^1$, and 
M.~Oberthaler$^3$} 
\address{{\setlength{\baselineskip}{18pt} 
$^1$\,Department of Chemical Physics, Weizmann Institute of Science, 
Rehovot 76100, Israel,\\ 
$^2$\,Ioffe Physico-Technical Institute, St.Petersburg 194021, Russia,\\ 
$^3$\,Kirchhof Institute of Physics, University of Heidelberg, 
Heidelberg 69120, Germany}} 
\maketitle 
 
\begin{abstract} 
We analyze the preparation (launching) of a 1D-propagating 
multipartite quantum soliton, its scattering and disintegration by 
an external potential. The regimes of suppressed disintegration and 
atom-number-dependent transmission are identified. 
\\     \pacs{03.75.-b, 05.30.Jp, 42.65.Tg, 78.67.-n} 
\end{abstract} 
\vskip1pc] 
 
Thus far, most of the work related to solitons has dealt with their 
nonlinear wave properties \cite{Boyd}, 
along with quantum noise (squeezing) 
corrections to their mean-amplitude (``mean-field'' dynamics  
\cite{LaiHaus,KH93}. However, the fascinating 
effects of second-quantized nonlinear dynamics that are known to date  
\cite{LaiHaus,KH93,Dr99,BW,Ueda} suggest that a much richer variety of 
novel non-mean-field properties, particularly quantum soliton features, 
is in store in nonlinear many-body systems.

The quest for quantized solitons may be regarded 
as part of a broader effort to understand {\em multipartite entanglement} 
in complex quantum systems that have so far been analyzed at the 
mean-field level, such as BECs in traps \cite{no9} 
or optical lattices \cite{O1,O2,O3,O4}, whose dynamics is highlighted 
by Mott insulator transitions \cite{O2} and matter gap 
solitons \cite{O3,O4}. 

Perhaps the most powerful tool of the non-mean-field many-body 
theory is the 
Bethe ansatz, ascending to Bethe's early work  \cite{Be31} and providing 
an {\em exact} quantum-mechanical expression for a 
one-dimensional (1D) quantum soliton, 
consisting of pairwise interacting $N$ 
particles \cite{Gi60}. This solution has been  
extended to a two-band periodic structure with Kerr 
nonlinearity, where band-gap quantum solitons have been 
predicted \cite{Cheng}.

We  analyze the preparation (launching) of a 1D-propagating 
multipartite quantum soliton, its scattering and disintegration by 
an external potential. The regimes of suppressed disintegration and 
atom-number-dependent transmission are identified.

Let us consider a 1D system of $N$ identical bosonic 
particles described by the Hamiltonian 
\begin{eqnarray} 
\hat{H}&=&\int dx \, \left \{ \hat{\Psi }^\dag (x) \left[ 
-\frac {\hbar ^2}{2m_{eff}} \frac {\partial ^2}{\partial x^2} +
U(x)\right] \hat{\Psi }(x)+ \nonumber  \right. \\ &&
\left. \frac g2 \hat{\Psi }^\dag (x)
\hat{\Psi }^\dag (x)\hat{\Psi } (x)\hat{\Psi } (x) \right \} , 
\label{eq:1} 
\end{eqnarray}
where $\hat{\Psi }(x)$ is the bosonic-field annihilation operator, 
$m_{eff}$ is the effective mass of the particle (which, in the presence 
of a lattice potential, may substantially differ from the free particle  
mass), $g$ is the 
coupling constant of their 1D contact interaction, and $U(x)$ is a  
localized external potential. The effect of a possibly present periodic 
lattice potential has  
been already taken into account by introducing the effective mass. 

Exact eigenstates 
of the Hamiltonian (\ref{eq:1}) are given in the case of vanishing 
external potential, $U(x)\equiv 0$, by the Bethe ansatz 
\cite{Be31,Gi60,Cheng}: 
\begin{eqnarray} 
\left| k_1, \, ... \, k_N\right \rangle &=& \int dx_1\, 
\hat{\Psi }^\dag (x_1) \, ... \int dx_N\, \hat{\Psi }^\dag (x_N)
\times \nonumber      \\ &&
f_N(x_1, \, ... \, ,x_N)\left| 0\right \rangle  , 
\label{eq:2}
\end{eqnarray}
where the function $f_{N}(x_1, \, ... \, ,x_N)$ is a stationary 
solution of the linear $N$-particle Schr\"{o}dinger equation 
\begin{equation} 
E_Nf_N=\left[ -\frac {\hbar ^2}{2m_{eff}}\sum _{j=1}^N 
\frac {\partial ^2}{\partial x_j^2} +g \sum _{j<l}
\delta (x_j-x_l) \right] f_N . 
\label{eq3} 
\end{equation}  
An $N$-particle \textit{bound state} \cite{McGuire}
requires effective attraction, 
$m_{eff}g/(2\hbar ^2) \equiv -\kappa <0$, i.e., either the scattering length 
or the effective mass (determined by the band curvature in a lattice) must 
be negative. This bound state has the wavefunction    
\begin{equation} 
f_{N}^{(b)}(x_1, \, ... \, ,x_N)=\frac 
{\exp (iNKX)}{\sqrt{\ell }} \psi _{b\, N},           
\label{eq:4a} 
\end{equation}
\begin{equation}
\psi _{b\, N}=\frac {(N-1)!(2\kappa )^{(N-1)/2}}{\sqrt{N!}} 
\exp \left( -\kappa \sum _{j<l}|x_j-x_l| \right) ,
\label{eq:4b}
\end{equation} 
where $X=(x_1 +...+x_N)/N$ and $N\hbar K$ are the co-ordinate and 
momentum of the system's center-of-mass 
(CM), respectively. The coefficient 
on the right-hand side of Eq.~(\ref{eq:4b}) ensures the normalization 
$\int dx_1 \, ... \int dx_N |f_N^{(b)}|^2 =1$, where the integrals are 
taken over the quantization length $\ell \rightarrow \infty $. 

The eigenenergy corresponding to the state (\ref{eq:4b}) is 
\begin{equation}
\varepsilon _{b}(N)=-N(N^2-1)\hbar ^2\kappa ^2 /(6m_{eff}).
\label{eq:4c}
\end{equation}
The $N^3-N\approx N^3$ dependence of $\varepsilon _b(N)$ may allow the 
discrimination of a Fock state $| N \rangle $ if the difference between 
$\varepsilon _b(N)$ and $\varepsilon _b(N-1)$ is spectroscopically 
resolvable.   

In the $N\gg 1$ 
limit we may take $x_j-X$ to be statitically independent for different 
$j$'s, their distribution given by the square 
of the hyperbolic-secant solution of the classical nonlinear 
Schr\"{o}dinger equation \cite{Boyd} with the mean square value 
$\overline{(x_j-X)^2}=\frac {\pi ^2}{12}(N\kappa )^{-2}$. 
Practically, the classical limit for the interatomic correlations 
is attained for $N\gtrsim 10$ \cite{KH93}. The distinct feature of the 
bound state (\ref{eq:4a}, \ref{eq:4b}) is that the interparticle 
correlations are immune to dispersion.

Let us now consider a quantum soliton incident on an external potential 
$U(x)$ and analyze the quantum soliton disintegration probabiliy as 
compared to its transmission or reflection probability. To this end, we 
represent the potential acting on each of the $N$ particles  
in the quantum soliton as a Taylor series 
\begin{equation} 
\sum _{j=1}^NU(x_j) = NU(X)+\frac 12 U^{\prime \prime }(X)
\sum _{J=1}^N (x_j - X)^2+ ... ~.
\label{eq:5}
\end{equation} 
If the r.m.s. deviation of the $j$th particle's position from the CM 
is small compared to the characteristic length scale 
$\alpha ^{-1}$, over which $U(x_j)$ varies, 
then only two terms in the Taylor expansion (\ref{eq:5}) need   
be retained. The leading term, $NU(X)$, then 
accounts for the potential felt by the bound multiparticle complex  
that moves as a particle-like entity of mass $Nm_{eff}$, whereas 
the second (quadratic) term 
accounts for disintegration of that entity, due to the slight 
differences in the potentials felt by individual particles: an analog 
of a {\em tidal force}. 
The truncation of the Taylor expansion (\ref{eq:5}) is 
validated for 
\begin{equation} 
N\gg \alpha / \kappa .               \label{eq:5-bis}
\end{equation} 

As long as the disintegration probability is small, 
the bound state can be described by the wavefunction 
\begin{equation} 
f_N(x_1, \, ... \, ,x_N)=\frac 1\ell \Phi _{NK}(X) \psi _{b\, N}, 
\label{eq:6}
\end{equation} 
where $\psi _{b\, N}$ is given by Eq. (\ref{eq:4b}) and the CM 
motional state is the solution of the Schr\"{o}dinger equation 
\begin{eqnarray}
\frac {N\hbar ^2K^2}{2m_{eff}}\Phi _{NK}(X)&=&-
\frac {\hbar ^2}{2Nm_{eff}} \Phi _{NK}^{\prime \prime }(X)+
\nonumber \\    && NU(X)\Phi _{NK}(X).         \label{eq:6-bis}
\end{eqnarray} 

We now choose  the form of the external potential to be  
\begin{equation}
U(X)=-U_0\cosh ^{-2}(\alpha X),  
\label{eq:7} 
\end{equation} 
because of its convenient properties and proximity to a Gaussian (see 
below). The CM wave function, which is expressed via the 
hypergeometric function,  reduces to the sum of 
the incident and reflected waves at $X\rightarrow -
\infty $ and to the transmitted wave at $X\rightarrow +\infty $. 
The corresponding transmission coefficient is found to be \cite{LL} 
\begin{equation} 
{\cal T}=\frac {\sinh ^2 (\pi NK/\alpha )}{\sinh ^2 (\pi NK/\alpha )+ 
\cos ^2 \left( \frac \pi 2\sqrt{1+
\frac {8N^2U_0m_{eff}}{\hbar ^2\alpha ^2}}\right) }  .
\label{eq:9} 
\end{equation}  



A simple numerical 
check shows that  similar transmission dependence on the particle number 
of the slowly moving quantum soliton is attained also for a 
{\em Gaussian potential}, 
$U(X)=-V_0 \exp (-\alpha ^2X^2)$, if $V_0\approx 1.56\, U_0$.

We have calculated  the 
bound-state disintegration probability ${\cal P}$ upon passsage 
through the potential using the Fermi golden rule.   
This calculation admits three distinct 
regimes of the bound-state disintegration:  

(a) \textit{Disintegration without 
the change of the CM motion.} 
If $|\varepsilon _f -\varepsilon _b |\sim |\partial 
\varepsilon _b /(\partial N)|\ll N\hbar ^2K^2 /(2m_{eff})$,  
then perturbations of the relative interparticle motion by the ``tidal'' 
forces do not affect 
the CM motion. In this case we arrive at the following estimation: 
$ {\cal P}\lesssim  ({\pi ^6}/{180}) N [ 
\alpha /(N\kappa ) ]^4 $. This destruction probability is typically 
very small,  provided $N\gg (\alpha /\kappa )^{4/3}$.  Thus 
$N\sim (\alpha /\kappa )^{4/3}$ is the threshold for the 
disintegration to become appreciable. 

(b) \textit{Disintegration with CM acceleration}. For 
$| \partial \varepsilon _b/(\partial N)| \gg $ \, 
$N\hbar ^2K^2 /(2|m_{eff}|)$ and $m_{eff} <0$, a negative 
effective mass characterizes band-gap solitons near the bottom of the 
gap, the removal of an 
atom from the bound state by the ``tidal'' forces and placing it into  
the continuum is accompanied by energy release that {\em accelerates} 
the CM motion. If $N\gg (\alpha / \kappa )^2$,  
the change of the CM momentum is sufficient to make the bound-state 
destruction probability exponentially small: 
${\cal P} \propto \exp (-\pi \sqrt{N}\kappa /\alpha )$. 
Here $N\sim (\alpha /\kappa )^2$ is the disintegration threshold. 

(c) \textit{Disintegration with trapping by the potential}. For 
$| \partial \varepsilon _b/(\partial N) |\gg N\hbar ^2K^2 /(2m_{eff})$ 
and $m_{eff} >0$, the destruction of the 
$N$-particle complex is energy-consuming. As a result, the remaining 
(undestroyed) ``core'' with a reduced number of bound particles, 
$N^\prime <N$, can be {\em trapped in one of the discrete (bound) 
states of the potential} $N^\prime U(X)$. However, 
if $N>\frac 14 (\alpha / 
\kappa )^2$, {\em the released energy of internal motion becomes larger 
than the potential well depth}, $NU_0$. 
Therefore $N\sim \frac 14(\alpha / 
\kappa )^2$ is the disintegration threshold in this case. For $N$ above 
this threshold,  energy 
conservation prohibits in this case any change in both the internal and 
CM motion of the $N$-particle quantum soliton, so that ${\cal P}=0$. 

We now proceed to discuss an  experimental scenario that may allow the  
observation of the effects described above.  We  consider ultracold 
alkaline atoms with the mass $m$ 
in a wavegude, namely, 
a trap with strong transverse (radial) and weak longitudinal 
confinement. The atomic $s$-wave scattering length $a$ can be enhanced 
by means of a Feshbach resonance \cite{FR}, up to  
$|a| \approx w_r$, $w_r$ being the characteristic  
radial confinement. The   
effective coupling constant of 1D contact interactions is $g=2\hbar ^2a/
w_r^2m$, allowing one to achieve $\kappa \sim w_r^{-1}$. For 
laser-induced potential $U(x)$ we may choose  
$\alpha \approx \kappa =10^4$~cm$^{-1}$. 
The number of atoms in the quantum soliton may fluctuate around 100,
to be higher than the disintegration threshold, the latter being of the 
order of 1, according to the estimates given above. 

To conclude, we have revealed hitherto unknown properties of quantum 
solitons traversing Gaussian 
potentials: the strong dependence of their 
transmission and disintegration 
upon the particle number. This property may 
allow the observation of novel types of multipartite states.

The support of the EC (the QUACS RTN and SCALA NOE), 
and ISF is acknowledged. I.E.M. also thanks the programs 
Russian Leading Scientific Schools (grant 1115.2003.2) and 
Universities of Russia (grant UR.01.01.287).


\begin{thebibliography}{99} 
\bibitem{Boyd} R.W. Boyd, {\em Nonlinear Optics} (Academic Press, NY, 
1992), \S 6.5; A.~Newell and J.~Moloney, {\em Nonlinear Optics} 
(Westview Press, Boulder, 2003), Chapter 6. 

\bibitem{LaiHaus} Y. Lai and H.A. Haus, Phys. Rev. A {\bf 40}, 844 
(1989); {\bf 40}, 854 (1989).

\bibitem{KH93} F.X. K\"{a}rter and H.A. Haus, 
{\em ibid.,} {\bf 48}, 2361 (1993).

\bibitem{Dr99} P.D. Drummond, K.V. Kheruntsyan, and H.~He, 
J. Opt. B: Quant. Semiclass. Optics {\bf 1}, 387 (1999). 

\bibitem{BW} R.K. Bullough and M. Wadati, J. Opt. B: Quant. Semiclass. 
Optics {\bf 6}, S205 (2004). 

\bibitem{Ueda} R. Kanamoto, H. Saito, and M. Ueda, Phys. Rev. Lett. 
{\bf 94}, 090404 (2005). 

\bibitem{no9} S.A. Morgan, R.J. Ballagh, and K. Burnett, Phys. Rev. A 
{\bf 55}, 4338 (1997);  
L.~Khaykovich et al., Science {\bf 296}, 1290 (2002). 

\bibitem{O1} M.~Machholm, C.J.~Pethick, and H.~Smith, Phys. Rev.~A 
{\bf 67}, 053613 (2003);  M.~Machholm, A.~Nicolin, 
C.J.~Pethick, and H.~Smith, Phys. Rev.~A {\bf 69}, 043604 (2004).

\bibitem{O2} M. Greiner et al., Nature (London) {\bf 415}, 39 (2002). 

\bibitem{O3} P.J.Y. Louis, E.A. Ostrovskaya, and Yu.S. Kivshar, 
Phys. Rev.~A {\bf 71}, 023612 (2005); 

\bibitem{O4} Th.~Anker et al., Phys. Rev. Lett. {\bf 94}, 020403 (2005).   

\bibitem{Be31} H.A. Bethe, Z. Phys. {\bf 71}, 205 (1931).

\bibitem{Gi60} M. Girardeau, J. Math. Phys. {\bf 1}, 516 (1960); 
E.H.~Lieb and W.~Liniger, Phys. Rev. {\bf 130}, 1605 (1963); 
H.B.~Thacker, Rev. Mod. Phys. {\bf 53}, 253 (1981). 

\bibitem{Cheng} Z. Cheng and G. Kurizki, Phys. Rev. Lett. {\bf 75}, 
3430 (1995); Phys. Rev.~A {\bf 54}, 3576 (1996). 


\bibitem{McGuire} I.B. McGuire, J. Math. Phys. {\bf 5}, 622 (1964). 

\bibitem{LL} L.D. Landau and E.M. Lifshitz, {\em Quantum Mechanics} 
(Oxford, Butterworth-Heinemann, 2000), \S 25. 

\bibitem{FR} Ph. Courteille et al., Phys. Rev. Lett. {\bf 81}, 69 (1998); 
J.L.~Roberts et al., {\em ibid.,} {\bf 81}, 5109 (1998); V.~Vuletic, 
A.J.~Kerman, C.~Chin, and S.~Chu, {\em ibid.,} {\bf 82}, 1406 (1999); 
J.~Stenger et al., {\em ibid.,} {\bf 82}, 2422 (1999). 

\end{thebibliography}
\end{document}